# The First Open-Source Framework for Learning Stability Certificates from Data

**Zhe Shen,** *IEEE Member*
**Stable Robotics Limited, UK**
**71-75 Shelton Street, Covent Garden**
**London, WC2H 9JQ**
**zheshen@ieee.org**

*Abstract*— **Before 2025, no open-source system existed that could learn Lyapunov stability certificates directly from noisy, real-world flight data. This work addresses that gap by proposing a data-driven approach that learns Lyapunov functions from trajectory data under realistic, noise-corrupted conditions. Unlike statistical anomaly detectors that only flag deviations, the proposed method assesses whether the system can still be certified as stable. Applied to public data from the 2024 SAS severe turbulence incident, this framework revealed that, within 60 seconds of the aircraft's descent becoming abnormal, no Lyapunov function could be constructed to certify system stability. To the best of our knowledge, this is also the first application of a data-driven Lyapunov-based stability verification method to real civil aviation data, achieved without any access to proprietary controller logic. The proposed framework is open-sourced and available at: https://github.com/HansOersted/stability**

## 1. INTRODUCTION

Establishing trust in control systems is a fundamental yet challenging task. While the stability of a controller is directly linked to the reliability of the system it governs. Many modern controllers operate as "black boxes," offering limited interpretability and no formal guarantees of stability under all conditions. This raises a critical question: how can stability be assessed when the internal dynamics of the controller remain unknown?

Indeed, establishing rigorous confidence in controllers has long remained a central challenge in automation, robotics, and aerospace. At the core of stability analysis lies Lyapunov theory [1], which, through its various extensions, remains the most widely accepted foundation for formal stability certification. A notable example is the concept of Input-to-State Stability (ISS), introduced by Sontag [2], which generalizes Lyapunov stability to systems subject to external disturbances or noise [3], thereby providing a natural framework for stability proofs in realistic conditions.

The significance of stability proofs has been deeply rooted in the design of controllers for robotics, aerospace vehicles, and other safety-critical systems [4]. In practice, controller design is often guided by the search for a stability certificate, most commonly a Lyapunov candidate, to ensure that the closed-loop system remains stable [5]. Such guarantees are essential: they certify that the system can reliably track the desired reference trajectory [6], even in the presence of bounded disturbances. By contrast, a controller without a stability proof is generally considered unreliable for use in high-precision or safety-critical applications.

Finding a stability proof, however, is often time-consuming and mathematically demanding, and in many cases may even be analytically intractable. For example, deriving a Lyapunov certificate for a linear controller is relatively straightforward and is often integrated directly into the design process [7]. In contrast, certifying the stability of relatively more sophisticated controllers, such as geometric control laws, requires advanced mathematical tools that can typically be accessible to specialized control theorists [8].

When controller design strictly follows the traditional paradigm focused on stability certificates—without accounting for unexpected malfunctions such as engine failures—a system may still exhibit instability in practice due to external noise. Such disturbances are rarely known beforehand in the design phase, and even a well-designed controller can become non-robust when exposed to real-world uncertainties, potentially invalidating the previously established Lyapunov candidate. To the best of our knowledge, no open-source framework has yet been demonstrated that can certify stability under realistic, noise-affected conditions in real time.

Pursuing real-time assurance of controller performance has become a recent focus, particularly with the development of data-driven methods. By leveraging the recent tracking history—such as time-stamped sequences of references and system states—it becomes possible to automatically construct stability certificates. This paradigm moves beyond model-dependent approaches, enabling real-time verification of black-box and data-driven controllers.

LYAPAK [9] is one of the commonly used MATLAB toolboxes for searching Lyapunov certificates. However, its admissible inputs exclude noise, meaning that stability verification becomes unreliable when applied to authentic trajectories contaminated by system disturbances. As a result, its applicability to real-world data remains limited. Similar approaches [10], [11] implemented in other programming languages or platforms share the same limitation: without explicit consideration of noise, the search for Lyapunov candidates can be significantly affected once disturbances are present in the data. As for numerical solvers, several methods impose a polynomial structure [12], [13] on the Lyapunov candidate, thereby restricting its functional form. Such constraints may rule out feasible certificates and, in turn, potentially excluding valid certificates.

To the best of the authors' knowledge, no open-source framework has been previously developed for learning stability certificates directly from real-world data, particularly in the presence of noise. This work introduces a data-driven approach that addresses this gap by automatically searching for Input-to-State Stability (ISS) certificates using neural networks.

To demonstrate the practical effectiveness of the proposed framework, it is evaluated on authentic flight data collected prior to the 2024 SAS severe turbulence incident [14]. By training Lyapunov candidates directly from the recorded trajectories, our method enables stability certification under extreme real-world conditions, illustrating its capability for real-time monitoring of controller integrity, in the presence of noise.

The remainder of this paper is organized as follows. Section 2 reviews the preliminaries of stability proofs. Section 3 formally states the problem, while Section 4 presents the proposed solution. The 2024 SAS flight incident was analyzed in Section 5 using the developed framework. Finally, Section 6 concludes the paper with discussions and future directions.

## 2. PRELIMINARY

The definition of stability in this work is Input-to-State Stability (ISS), introduced by Sontag [2]. ISS characterizes how the system remains bounded (stable) in the presence of the external disturbances. In contrast to conventional Lyapunov stability, ISS allows small deviations from the equilibrium state, e.g., the state disturbance introduced by the noise.

ISS can be interpreted as the existence of a positive-definite Lyapunov candidate $V$ [2] satisfying

$$\dot{V}(\boldsymbol{x}) \leq -\alpha(\|\boldsymbol{x}\|) + \gamma(\|\boldsymbol{u}\|), \tag{1}$$

where $\boldsymbol{x}$ is the state, $\boldsymbol{u}$ is the input, $\alpha(\cdot)$ and $\gamma(\cdot)$ are class-$\mathcal{K}$ functions.

In this work, ISS is simplified for practical data-driven analysis. More specifically, a positive-definite Lyapunov candidate, $V$, is demanded, meeting the following criteria:

$$\dot{V} \leq \varepsilon, \tag{2}$$

where $V(\mathbf{0}) = 0$, and $\varepsilon$ is a positive finite constant, also waiting to be found.

It is worth noting that the existence of a Lyapunov candidate provides a constructive proof of stability. However, the converse is not necessarily true: failing to identify a Lyapunov candidate does not, in itself, imply that the system is unstable without further analysis.

In this research, inspired by the previous research [15], the Lyapunov candidate in Formula (2) is structured in the form below:

$$V(\boldsymbol{\xi}) = \boldsymbol{\xi}^\top Q \boldsymbol{\xi}, \tag{3}$$

where $\boldsymbol{\xi} \in \mathbb{R}^n$ represents the state-related term, and $Q \in \mathbb{R}^{n \times n}$ is a symmetric positive definite constant matrix.

This designation in Formula (3) naturally guarantees the positive-definite property demanded in the Lyapunov candidate in Formula (2).

## 3. PROBLEM STATEMENT

Consider an unknown dynamical system, where "unknown" refers to the fact that the governing differential equations are not explicitly available. The only accessible information consists of the observed state trajectories, $\boldsymbol{x}(t)$, sampled over time, which may also be contaminated by measurement noise.

In addition, the reference trajectory, $\boldsymbol{r}(t)$, which the system is intended to track, is also available as a time-dependent signal.

Define the tracking error, $\boldsymbol{e}(t)$, below:

$$\boldsymbol{e}(t) = \boldsymbol{r}(t) - \boldsymbol{x}(t). \tag{4}$$

Higher-order derivatives of the tracking error, such as $\dot{\boldsymbol{e}}(t)$, $\ddot{\boldsymbol{e}}(t)$, and so forth, from Formula (4), can be obtained using numerical differentiation methods when required.

The problem can thus be stated as follows: find a Lyapunov candidate $V$, in the form given by (3), that fulfills the Input-to-State Stability (ISS) criterion specified in (2).

## 4. DATA-DRIVEN STABILITY CERTIFICATE

This section presents the proposed solution by introducing the designed structure of the Lyapunov candidate, the associated loss function for neural network training, the neural network architecture used to identify the parameters of this structure, and the estimation of the noise term $\varepsilon$ in Formula (2).

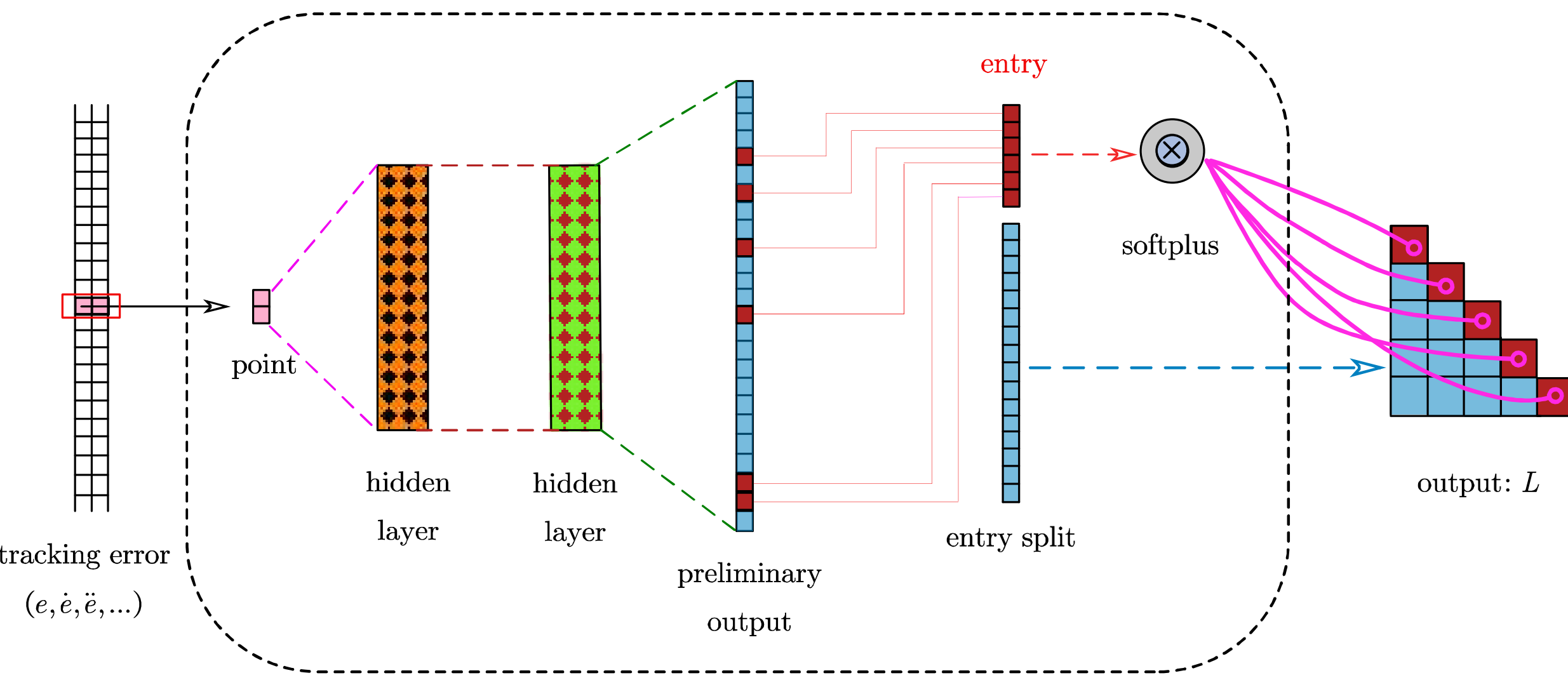

**Figure 1.** Neural network for learning the Cholesky factor. The inputs are the tracking error and its derivatives, and the output is the estimated Cholesky factor, $\boldsymbol{L}$, which constitutes the parameter $\boldsymbol{Q} = \boldsymbol{L}\boldsymbol{L}^{\top}$ used in the Lyapunov candidate (3).

*Lyapunov Candidate Structure*

To guarantee the positive-definite property of the Lyapunov function, the parameter $Q$ in (3) is factorized using the Cholesky decomposition:

$$Q = LL^{\top}, \tag{5}$$

where $L$ is the Cholesky factor [16] which is a real lower triangular matrix with positive diagonal entries.

Once the Cholesky factor, $L$, is determined, the Lyapunov candidate is subsequently obtained based on Formula (3) and (5), using the state-related term, $\boldsymbol{\xi}$, defined as

$$\boldsymbol{\xi} = \begin{bmatrix} e \\ \dot{e} \end{bmatrix}, \tag{6}$$

where $e$ represents the tracking error defined in Formula (4).

The subsequent subsections present the methods for learning the Cholesky factor, $L$, using a neural network.

*Loss Function*

Before introducing the architecture of the neural network, we define the following loss function for its training process. The auxiliary function is given by

$$h(\boldsymbol{\xi}) = \dot{V}(\boldsymbol{\xi}) + \gamma, \tag{7}$$

where $\gamma > 0$ is a small positive constant and $\dot{V}(\boldsymbol{\xi})$ is computed according to (3), (5), and (6).

Based on this, the loss function is defined as

$$\psi(\boldsymbol{\xi}) = \max\{0, h(\boldsymbol{\xi})\}. \tag{8}$$

Intuitively, driving $\psi(\boldsymbol{\xi})$ toward zero enforces $\dot{V}(\boldsymbol{\xi}) \leq 0$ is negative, thereby satisfying the ISS condition in (2).

The neural network introduced in the next subsection aims to minimize this loss by updating its weights and biases through gradient-based training at each epoch.

*Neural Network Architecture*

The purpose of the proposed neural network is to identify the Cholesky factor in (5). With $L$ determined, the Lyapunov candidate is directly constructed according to (3) and (5). The adopted architecture, consistent with the parallel work in [17], is depicted in Fig. 1.

The tracking error defined in (4) is passed through hidden layers, producing a preliminary output whose elements are unconstrained, i.e., they may take either positive or negative values.

Since the entries of the Cholesky factor $L$ in (5) are required to be strictly positive, the preliminary output is split into two groups.

Elements in the first group are directly used as the non-diagonal entries of the real lower-triangular matrix.

Elements in the second group are passed through the **Softplus** activation function to enforce positivity, and the resulting values are adopted as the diagonal entries of the Cholesky factor.

Appropriate activation functions are used to prevent numerical issues in training. At each epoch, the weights and biases are updated via gradient descent; gradients are then reset, and the updated weights serve as the initialization for the next epoch.

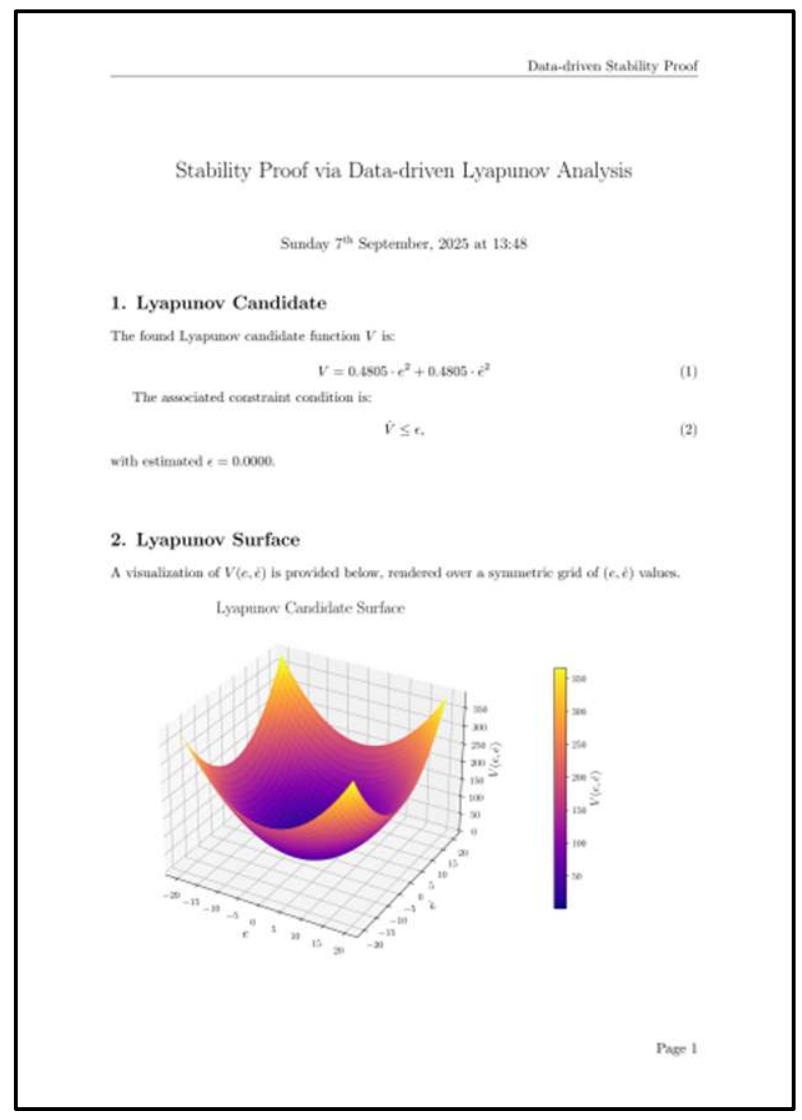
Data-driven Stability Proof

Stability Proof via Data-driven Lyapunov Analysis

Sunday 7th September, 2025 at 13:48

**1. Lyapunov Candidate**

The found Lyapunov candidate function $V$ is:

$$V = 0.4805 \cdot e^2 + 0.4805 \cdot \dot{e}^2 \quad (1)$$

The associated constraint condition is:

$$\dot{V} \leq \epsilon, \quad (2)$$

with estimated $\epsilon = 0.0000$.

**2. Lyapunov Surface**

A visualization of $V(e, \dot{e})$ is provided below, rendered over a symmetric grid of $(e, \dot{e})$ values.

Page 1

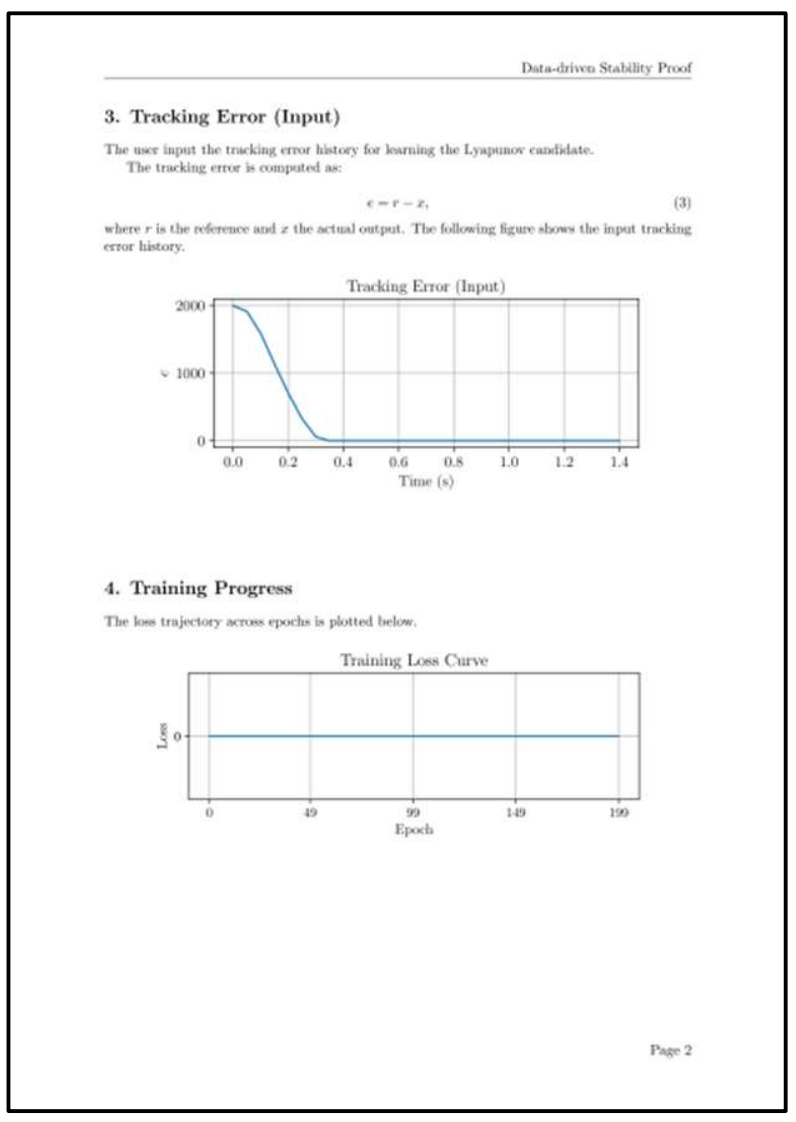
Data-driven Stability Proof

**3. Tracking Error (Input)**

The user input the tracking error history for learning the Lyapunov candidate.

The tracking error is computed as:

$$e = r - x, \quad (3)$$

where $r$ is the reference and $x$ the actual output. The following figure shows the input tracking error history.

**4. Training Progress**

The loss trajectory across epochs is plotted below.

Page 2

**Figure 2.** Sample stability analysis report automatically generated.

For further implementation details, the open-source code is available at https://github.com/HansOersted/stability.

*Estimation of the Noise*

Note that in real datasets, noise is inevitably present, which often leads to a positive loss function and thus violates the condition $\dot{V}(\boldsymbol{\xi}) \leq 0$. To account for this, the noise level is estimated in the final step.

Specifically, unexpected noise may spoil the condition $\dot{V}(\boldsymbol{\xi}) \leq 0$ at certain time points. We therefore define the smallest non-negative $\varepsilon$, such that

$$\dot{V}(\boldsymbol{\xi}) \leq \varepsilon \quad (9)$$

for all sampled tracking errors $\boldsymbol{\xi}$ fed into the neural network.

It is worth noting that noise is not the only source of violation of the condition $\dot{V}(\boldsymbol{\xi}) \leq 0$. The nonlinear behavior of the Lyapunov candidate may also contribute to such violations. Both effects are collectively captured in the noise-related parameter $\varepsilon$.

*Expected Output*

To support reproducibility and practical application, the open-source Python implementation automatically generates a stability analysis report in PDF format. The report includes the identified Lyapunov candidate, the corresponding Lyapunov surface, the input tracking error, and the *Training Loss* history. A sample report is shown in Fig. 2.

*Training Loss*

To visualize the violation of the ISS condition during the learning process, define the *Training Loss* as

$$\dot{V}(\boldsymbol{\xi}), \quad (10)$$

which serves as a measure of the violation of Formula (2), given that $\varepsilon = 0$.

Interestingly, when the *Training Loss* is zero, as observed in the sample report, it indicates that the initial weights and biases of the neural network already satisfy $\psi(\boldsymbol{\xi}) = 0$ in Formula (8), meaning no further updates are needed during training..

In this case, the Lyapunov candidate corresponding to the initial weights and biases already meets the ISS condition. This scenario often occurs in stable systems with multiple Lyapunov candidates, for example, due to scaling effects in a linear Lyapunov candidate.

## 5. Case Study: 2024 SAS Flight Incident

To demonstrate the applicability of the proposed framework, the altitude history from the 2024 SAS incident [14] is analyzed. The data inherently contain turbulence-induced disturbances—some recoverable by the controller, others not. The framework is applied to extract stability proofs, thereby demonstrating its performance.

*Incident Background*

In March 2024, Scandinavian Airlines (SAS) reported a severe turbulence incident on a transatlantic flight [14], resulting in an emergency engine shutdown and a temporary loss of altitude.

The event highlighted the importance of evaluating the robustness of autopilot and flight control systems under extreme atmospheric disturbances. For a stability verification perspective, such real-world disturbances constitute a highly relevant and challenging test case.

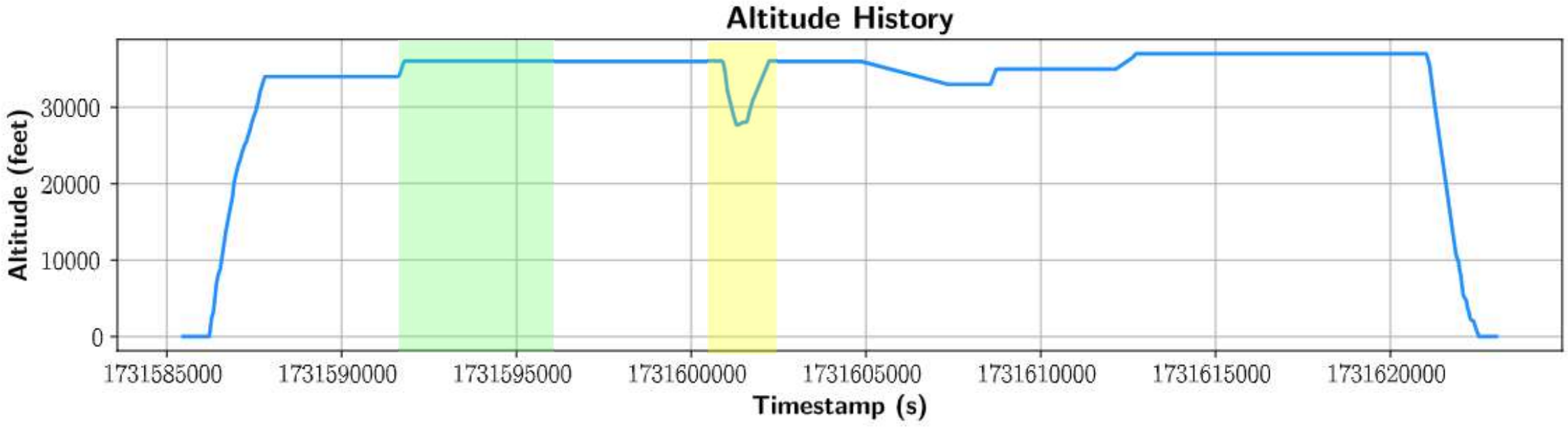

**Figure 3.** Altitude history of the SAS flight involved in the 2024 incident.

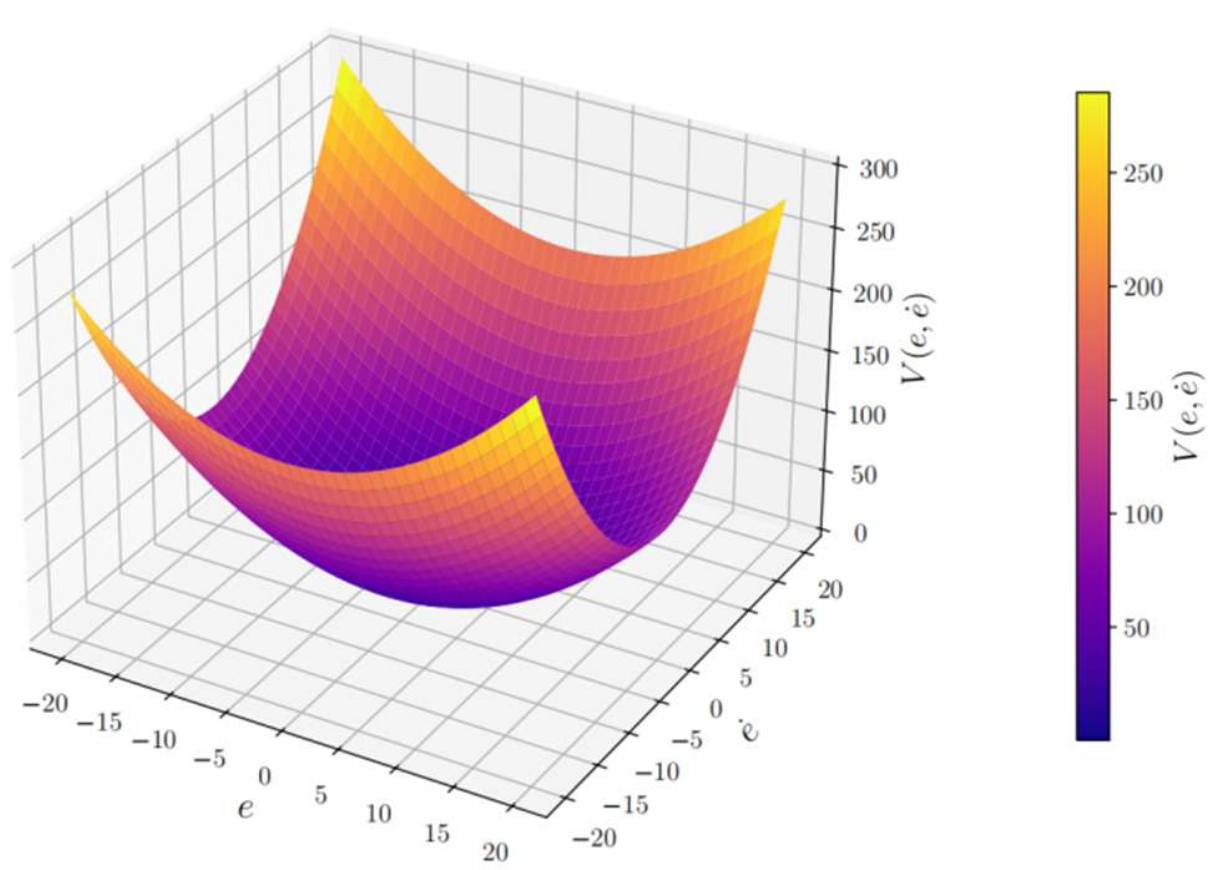

**Figure 4.** Lyapunov candidate surface corresponding to the beginning of the SAS flight involved in the 2024 incident.

*Flight History*

The complete flight profile containing the incident is shown in Fig. 3.

The time window highlighted in yellow in Fig. 3. corresponds to the moment when the aircraft, flying over Greenland, experienced a sudden altitude loss attributed to a control system anomaly. In such situations, pilots are required to take manual control if the autopilot is no longer capable of stabilizing the aircraft.

However, it is extremely challenging to determine at the right moment whether the controller has lost its stabilizing capability. To the best of our knowledge, no prior open-source framework has addressed this challenge by learning a stability certificate directly from real operational data.

The ability to provide real-time information about the controller's stability is crucial for making timely judgments: if the controller is deemed incapable of recovery, further corrective actions can be initiated promptly before further degradation of system performance.

In this study, two datasets of interest were selected for training. The first time window, highlighted in green, corresponds to the normal ascent phase and subsequent altitude maintenance, during which noise was introduced. The second time window, highlighted in yellow, covers the incident itself, characterized by the abnormal loss of altitude.

The selected flight data were first resampled at a rate of 30 s before computing the first and second derivatives. These processed signals were then used as training inputs to the neural network described in Section 4.

The results of training for stability proofs are presented in the following subsection.

*Data-driven Stability Proof*

The first highlighted dataset yields the learned Lyapunov candidate

$$V = 0.2425 \cdot e^2 - 0.0268 \cdot e \cdot \dot{e} + 0.4804 \cdot \dot{e}^2, \quad (11)$$

with the associated constraint condition

$$\dot{V} \leq \varepsilon, \quad (12)$$

where the noise-related constant $\varepsilon = 4.8871$.

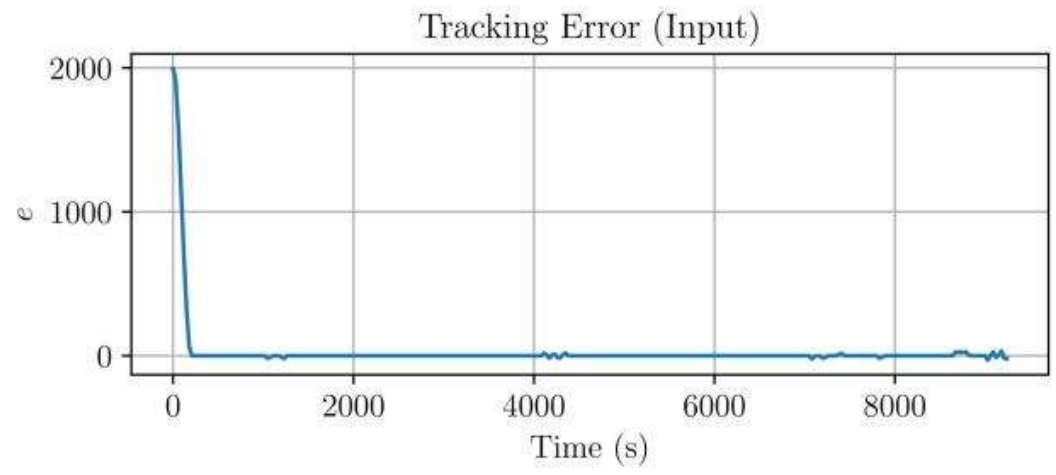

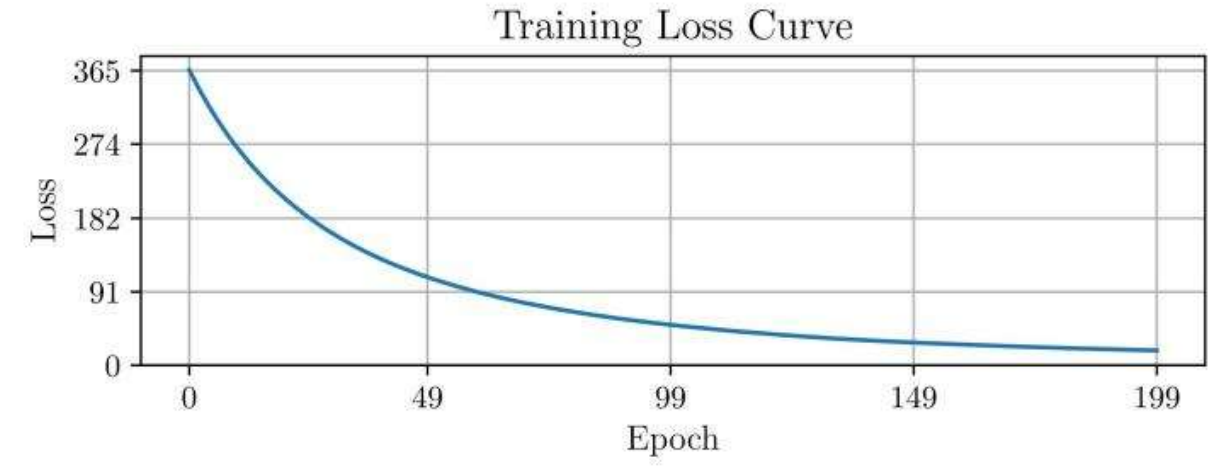

**Figure 5.** Tracking history time window till the onset of the incident and the resulting *Training Loss***.**

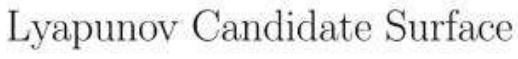

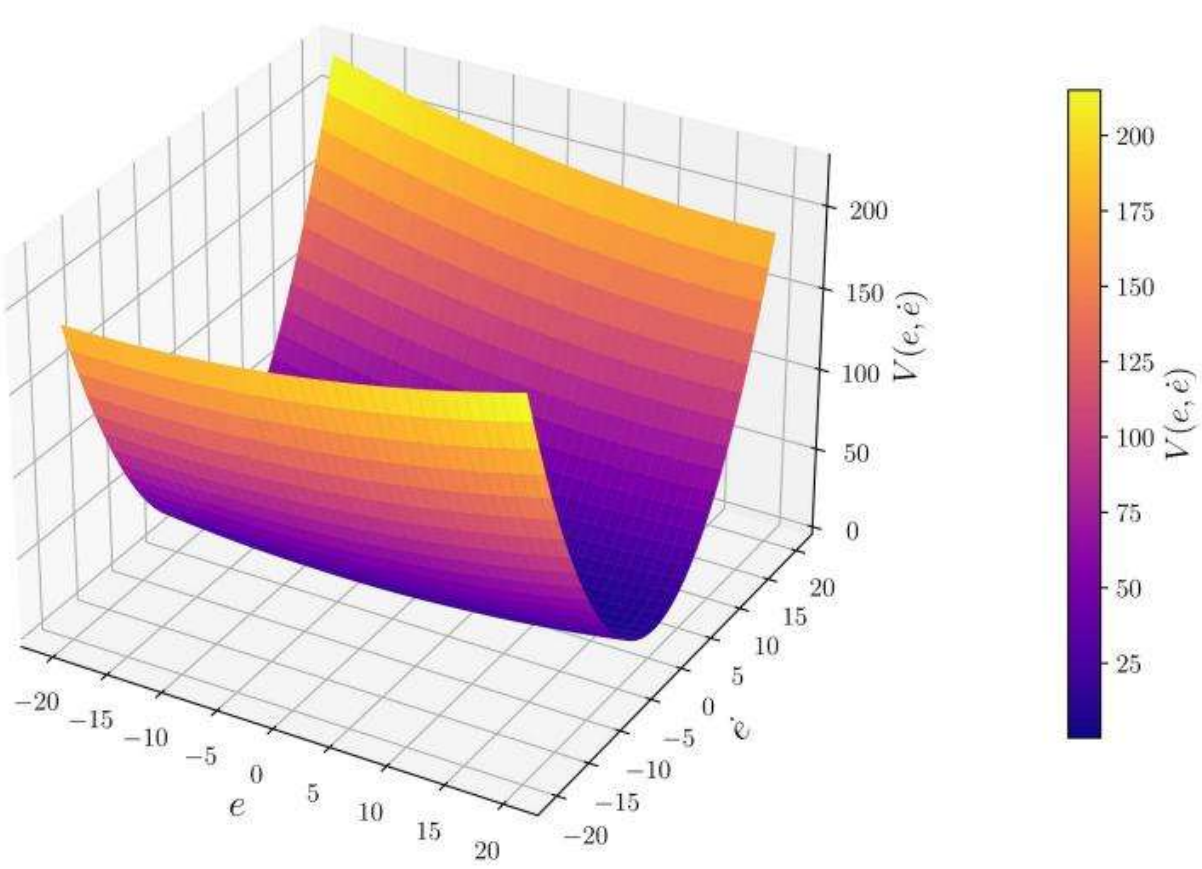

**Figure 6.** Lyapunov surface corresponding to the time window till the onset of the incident.

The results in (11) and (12) provide a concrete Input-to-State Stability (ISS) criterion as defined in (2), allowing us to conclude that the aircraft remained ISS stable during this time window. The corresponding Lyapunov surface is plotted in Fig. 4.

Importantly, this experiment demonstrates, to the best of our knowledge, one of the first applications of a neural network for learning a stability certificate directly from noisy real-world data.

For the second highlighted dataset corresponding to the incident, the neural network failed to identify a stability certificate; instead, the weights diverged during the training process. This outcome does not necessarily imply that the system was unstable but rather indicates that no Lyapunov candidate satisfying the given conditions was identified under the employed procedure.

*Lyapunov Candidate at the Onset of the Incident*

In training the altitude history at the beginning of the incident, corresponding to the tracking error in Fig. 5, the framework successfully identifies the stability proof.

The corresponding *Training Loss* history, which is defined in Formula (10), is also included in Fig. 5, and the resulting Lyapunov candidate (with $\varepsilon = 2.3547$), is given by

$$V = 0.0253 \cdot e^2 - 0.0479 \cdot e \cdot \dot{e} + 0.4919 \cdot \dot{e}^2, \quad (13)$$

which corresponds to the Lyapunov surface in Fig. 6.

It can be observed that the Lyapunov surface reshapes and grows more slowly along certain directions, compared to the healthy case with a functional controller.

## 6. CONCLUSION

This paper presented an open-source framework for learning Lyapunov stability certificates directly from noisy, real-world data.

Unlike traditional methods that require analytical models and handcrafted proofs, the proposed method employs neural networks and Cholesky factorization to automatically construct Input-to-State Stability (ISS) certificates from trajectory data.

The framework was validated using flight data from the 2024 SAS turbulence incident. During the normal flight window, a valid Lyapunov candidate was successfully obtained, yielding a concrete ISS certificate.

In contrast, during the incident window, the neural network failed to produce a stability certificate, which is consistent with expectations since a stability proof cannot be established for an unstable system. However, the absence of a certificate does not necessarily imply instability, but rather indicates that no Lyapunov function satisfying the given conditions was identified.

To the best of our knowledge, this work represents one of the earliest open-source studies on learning stability certificates from authentic, noisy flight data.

## Code and Contribution

The code in this research is open source and available at: https://github.com/HansOersted/stability

## BIOGRAPHY

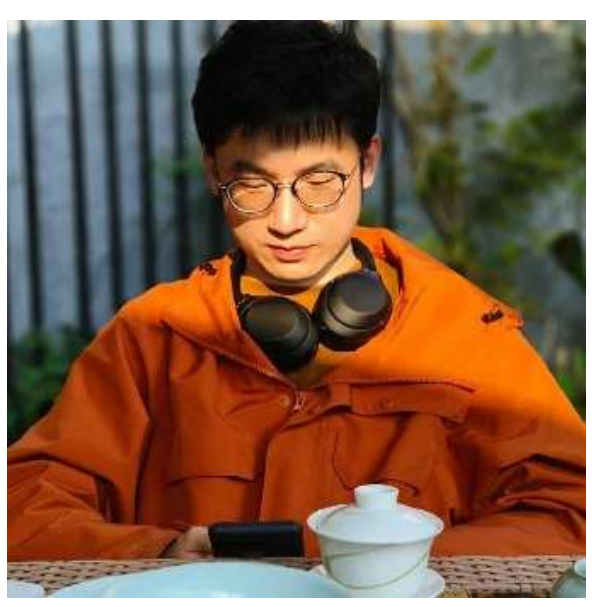

***Zhe Shen*** *received the M.Sc. degree in Automation and Robotics from the Technical University of Denmark in 2019, and the Ph.D. degree in Aeronautics and Astronautics from The University of Tokyo, Japan, in 2023. From 2024 to August 2025, he was a Postdoctoral Researcher with the University of Southern Denmark, where he contributed to the EUROfusion project. Since September 2025, he has been the founder of Stable Robotics Limited., United Kingdom, focusing on controller quality monitoring and risk assessment for applications in robotics, aviation, and nautical systems.*